\documentclass[aps,prb,preprint,superscriptaddress,floats]{revtex4}
\usepackage{txfonts}
\usepackage{amssymb}
\usepackage{graphicx}
\usepackage[rightcaption]{sidecap}
\usepackage{ulem}
\usepackage{color}

\begin{document}


\title{Marginal-Fermi-Liquid-like Behavior without Pseudogap in Infinite-Layer Nickelates}

\author{Yu Fan}
\affiliation{Advanced Materials Laboratory, State Key Laboratory of Surface Physics, and Department of Physics, Fudan University, Shanghai 200433, China}

\author{Zhitong An}
\affiliation{Advanced Materials Laboratory, State Key Laboratory of Surface Physics, and Department of Physics, Fudan University, Shanghai 200433, China}

\author{Xiang Ding}
\affiliation{Advanced Materials Laboratory, State Key Laboratory of Surface Physics, and Department of Physics, Fudan University, Shanghai 200433, China}

\author{Xingtian Sun}
\affiliation{Advanced Materials Laboratory, State Key Laboratory of Surface Physics, and Department of Physics, Fudan University, Shanghai 200433, China}

\author{Yutong Chen}
\affiliation{Advanced Materials Laboratory, State Key Laboratory of Surface Physics, and Department of Physics, Fudan University, Shanghai 200433, China}

\author{Zhihui Chen}
\affiliation{Advanced Materials Laboratory, State Key Laboratory of Surface Physics, and Department of Physics, Fudan University, Shanghai 200433, China}

\author{Shenglin Tang}
\affiliation{Advanced Materials Laboratory, State Key Laboratory of Surface Physics, and Department of Physics, Fudan University, Shanghai 200433, China}

\author{Chihao Li}
\affiliation{Advanced Materials Laboratory, State Key Laboratory of Surface Physics, and Department of Physics, Fudan University, Shanghai 200433, China}

\author{Jiahao Ye}
\affiliation{Advanced Materials Laboratory, State Key Laboratory of Surface Physics, and Department of Physics, Fudan University, Shanghai 200433, China}

\author{Timur Kim}
\affiliation{Diamond Light Source Ltd., Harwell Science and Innovation Campus, Didcot, OX11 0DE, United Kingdom}

\author{Haichao Xu}
\email{xuhaichao@fudan.edu.cn}
\affiliation{Advanced Materials Laboratory, State Key Laboratory of Surface Physics, and Department of Physics, Fudan University, Shanghai 200433, China}
\affiliation{Shanghai Research Center for Quantum Sciences, Shanghai 201315, China}

\author{Rui Peng}
\email{pengrui@fudan.edu.cn}
\affiliation{Advanced Materials Laboratory, State Key Laboratory of Surface Physics, and Department of Physics, Fudan University, Shanghai 200433, China}
\affiliation{Shanghai Research Center for Quantum Sciences, Shanghai 201315, China}

\author{Donglai Feng}
\email{dlfeng@hfnl.cn}
\affiliation{New Cornerstone Science Laboratory, Hefei National Laboratory, Hefei 230026, China}

\date{\today}

\begin{abstract}
Pseudogap formation, strange-metal behavior and unconventional superconductivity are closely intertwined in hole-doped cuprates, yet their relationship remains unresolved.
Infinite-layer nickelates offer a distinct $3d^9$-derived platform to address this question by combining a cuprate-like Ni $d_{x^2-y^2}$ Fermi surface with multiband electronic degrees of freedom.
Here we use angle-resolved photoemission spectroscopy to resolve the low-energy spectral function of superconducting La\(_{0.8}\)Ca\(_{0.2}\)NiO\(_2\) and parent LaNiO\(_2\) thin films. In La\(_{0.8}\)Ca\(_{0.2}\)NiO\(_2\), the electronic self-energy Im\(\Sigma(\omega)\) is approximately linear in energy and its slope increases from \((\pi/2,\pi/2)\) to \((\pi,0)\), revealing momentum-dependent marginal-Fermi-liquid-like scattering. Both films show a progressive suppression of low-energy spectral weight from the diagonal direction toward \((\pi,0)\), with stronger suppression in parent LaNiO\(_2\). 
However, finite Fermi-level spectral weight persists around the entire Fermi surface, with no leading-edge shift or back-bending indicative of pseudogap formation in either the electron pocket or the cuprate-like hole band. Our results demonstrate that momentum-selective correlations and marginal-Fermi-liquid-like scattering can occur without a detectable cuprate-like pseudogap, providing a benchmark for identifying the essential normal-state electronic ingredients of high-temperature superconductivity.
\end{abstract}

\maketitle

In strongly correlated electron systems, the breakdown of Landau Fermi-liquid theory often gives rise to emergent quantum states. A canonical example is provided by hole-doped cuprates, which are derived from a parent $3d^9$ electronic configuration and exhibit a rich phase diagram featuring pseudogap behavior \cite{Normannature1998,DamascelliRMP2003,LoeserScience1996,VishikPNAS2012}, strange-metal transport \cite{VallaScience1999,LegrosNatPhys2019}, and high-temperature superconductivity \cite{BednorzZPhysB1986}. The pseudogap is characterized by a momentum-dependent depletion of low-energy spectral weight above the superconducting transition temperature~\cite{Normannature1998,DamascelliRMP2003,LoeserScience1996,VishikPNAS2012,HashimotoNatPhys2010}, whereas the strange-metal state exhibits linear-in-temperature resistivity and anomalous quasiparticle dynamics that are inconsistent with conventional Fermi-liquid theory, interpreted in terms of a marginal Fermi-liquid~\cite{VallaScience1999,LegrosNatPhys2019}. 
These phenomena are central to the phenomenology of hole-doped cuprates and are often discussed in connection with high-temperature superconductivity~\cite{VarmaRPP2016,Emerynature1995}. However, because pseudogap behavior, strange-metal phenomenology and high-temperature superconductivity are entangled in the cuprate family, a key open question is whether their coexistence reflects a universal principle of hole-doped Mott systems or instead depends on material-specific aspects of the underlying electronic structure.

Infinite-layer nickelates provide a unique opportunity to address this question. They retain key electronic ingredients of cuprates, including a nominal $3d^9$ electronic configuration and a large $d_{x^2-y^2}$-derived Fermi surface \cite{BotanaPRX2020}. At the same time, they exhibit important distinctions, including an additional three-dimensional electride-like $\beta$ band \cite{dingNSR2024}, multiband superconductivity \cite{HeptingNatMater2020}, and a markedly different charge-transfer energy scale \cite{JiangPRL2020,LeePRB2004}. Recent transport measurements on high-quality infinite-layer nickelate films have revealed linear-in-$T$ resistivity reminiscent of the strange-metal behavior in cuprates~\cite{leenature2023}. Whether other hallmark phenomena of cuprates, particularly pseudogap formation and marginal-Fermi-liquid behavior, are also present remains an open question.
Momentum-resolved measurements of the low-energy quasiparticle features are therefore essential for establishing the nature of the normal state in infinite-layer nickelates and for clarifying how pseudogap, strange-metal and superconducting phenomena evolve in a distinct correlated-electron platform.

Direct determination of the momentum-resolved quasiparticle features, however, is particularly challenging in infinite-layer nickelates. Due to the poor surface quality of infinite-layer nickelates after topotactic reduction, angle-resolved photoemission spectroscopy (ARPES)
studies have been limited to large-scale band structure measurements \cite{dingNSR2024,SunSciAdv2025,LiPRL2025}, leaving the low-energy spectral function unresolved. In this work, we overcome these limitations by fabricating 
infinite-layer nickelate thin films with state-of-the-art quality, enabling spectral weight analysis near the Fermi level by ARPES measurements. 
\par
\par

\begin{figure*}[b]
    \centering
    \includegraphics[width=170mm]{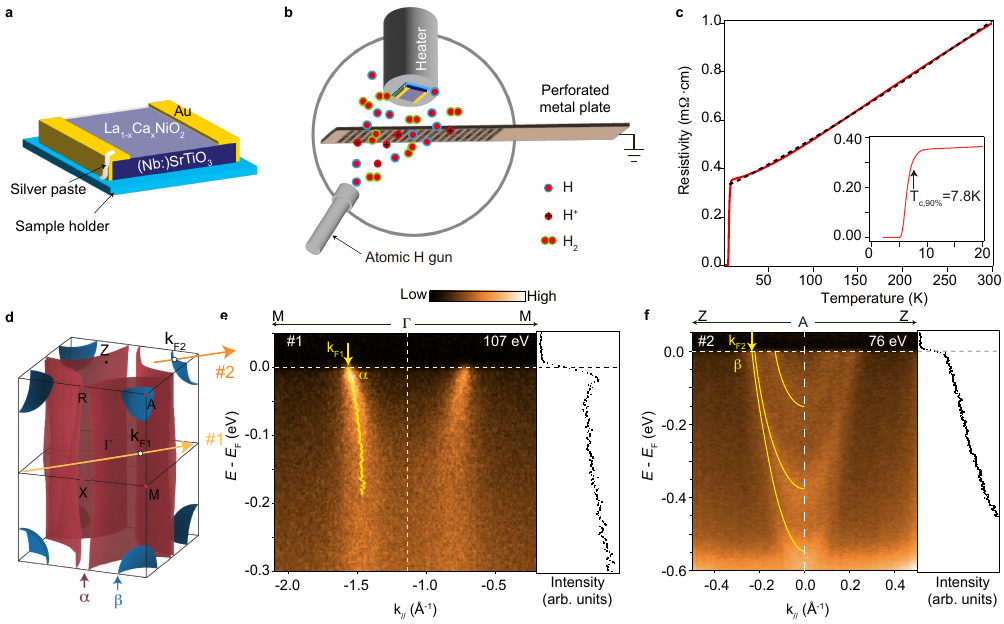}  
    \caption{\textbf{Improved surface quality enabling observation of quasiparticle peak in ARPES studies on La$_{0.8}$Ca$_{0.2}$NiO$_2$.}
\textbf{a,b}, Schematic illustrations of surface grounding of the heterostructures for ARPES measurements (\textbf{a}) and the atomic-hydrogen reduction setup using a perforated mask (\textbf{b}). \textbf{c}, Temperature-dependent resistivity showing a superconducting transition with $T_{\mathrm{c},90\%}=7.8$~K.
The dashed line is a power-law relation R(T)=$cT$$^\alpha$+$R_0$ ($\alpha = 1.14$) fit to the high-temperature data.
All subsequent ARPES measurements were performed at the normal state of 10~K. 
\textbf{d}, Schematic three-dimensional Brillouin zone and Fermi-surfaces of La$_{1-x}$Ca$_x$NiO$_2$ \cite{dingNSR2024}. The momentum cuts used in panels e and f are indicated.
\textbf{e}, Photoemission spectrum measured at $h\nu=107$~eV along the $M$--$\Gamma$--$M$ direction, showing the dispersion of $\alpha$ band. The yellow open circles mark the band dispersion extracted from the momentum distribution curve (MDC) fitting, and the corresponding energy distribution curve (EDC) at the Fermi momentum $k_{F1}$ is shown on the right panel. 
\textbf{f}, Same as panel (\textbf{e}) but measured at $h\nu=76$~eV along $Z$-$A$-$Z$ direction, showing the dispersion of $\beta$ band and its quantum well states \cite{LiPRL2025}. The yellow curves mark the band dispersion extracted from MDCs and the second derivative of the spectrum (Supplementary Fig.S5). The corresponding EDC at $k_{F2}$ is shown on the right panel.
}

    \label{fig:figure1}
\end{figure*}

\vspace{1em}

\noindent\textbf{Surface-quality optimization for quasiparticle-resolved ARPES}

To ensure reliable ARPES gap measurements free from the photoemission charging effect\cite{SohnNC2021}, gold was deposited on the edges and side facets of the (Nb:)SrTiO$_3$ substrates and electrically connected to the sample holder using silver paste (Fig.~1a). 
21-unit-cell-thick perovskite thin films were grown on (Nb:)SrTiO$_3$ by oxide molecular beam epitaxy, and then reduced to infinite-layer La$_{0.8}$Ca$_{0.2}$NiO$_2$ and LaNiO$_2$ by $\textit{in-situ}$ atomic hydrogen. 
A central challenge in ARPES studies is the trade-off between preserving surface crystallinity and achieving sufficient topotactic reduction to infinite-layer phase~\cite{kawaiAPL2009,leeAPLMater2020,OnozukaDalton2016}.
To address this issue, we previously employed a protective mask during hydrogen-atom reduction to minimize surface damage and enable ARPES measurements\cite{dingNSR2024}. Here, a modified mask with arrays of fine apertures (Fig.~1b) further improves the surface quality by promoting uniform hydrogen diffusion while suppressing direct H$^+$ bombardment. The reduced films show sharp X-ray diffraction (XRD) peaks characteristic of the infinite-layer phase (Supplementary Fig.~S1),
indicative of successful reduction and good crystallinity. Transport measurements on optimally doped films reveal a superconducting transition at approximately 7.8 K
(Fig.~1c), consistent with previous reports on La$_{0.8}$Ca$_{0.2}$NiO$_2$~(refs. \onlinecite{Chowarxiv2022,BalakrishnanNatCommun2024,ZengSciAdv2022}). The normal-state resistivity deviates from the quadratic temperature dependence expected for a Fermi liquid and instead follows a nearly linear-in-$T$ form, with $\alpha$ = 1.14 for the sample shown in Fig.~1c and a value closer to unity in a higher-quality sample, reminiscent of strange-metal-like transport (Supplementary Fig.~S2).

With improved surface crystallinity, the dispersive spectral weight  relative to the background is progressively enhanced (Supplementary Fig.~S3). The spectral weight near the Fermi level is increased, and momentum
distribution curve (MDC) linewidth is narrowed by approximately a factor of two, suggesting an increased quasiparticle lifetime with reduced disorder.  
In the optimized films, well-defined quasiparticle peaks are observed for the first time in this material system (Fig.~1e), emerging from the $d_{x^2-y^2}$-orbital-derived $\alpha$ band near $(\pi/2,\pi/2)$ (ref.\onlinecite{dingNSR2024}). 
Note that we use the term quasiparticle in a spectroscopic sense, referring to dispersive peak-like features in the single-particle spectral function which can be observed in a strange metal state \cite{ChangNatCommun2013,PlatePRL2005}, rather than long-lived Landau quasiparticles. 
The resulting spectral quality is comparable to that of early cuprate ARPES studies \cite{DamascelliRMP2003}, enabling quantitative analysis of the low-energy electronic structure.
The interstitial-$s$-dominated $\beta$ band \cite{LiPRL2025} exhibits much weaker electronic correlations than the $\alpha$ band (Supplementary Fig.S4), qualitatively consistent with the previous report on (La,Sr)NiO$_2$ \cite{SunSciAdv2025}. Its large Fermi velocity results in a less pronounced quasiparticle peak, while the well-resolved quantum-well states (Fig.~1f, see Supplementary Fig.S5 for further analysis) further demonstrate the high crystalline and surface quality of the films \cite{LiPRL2025}.
\begin{figure*}[htbp]
    \centering
    \includegraphics[width=170mm]{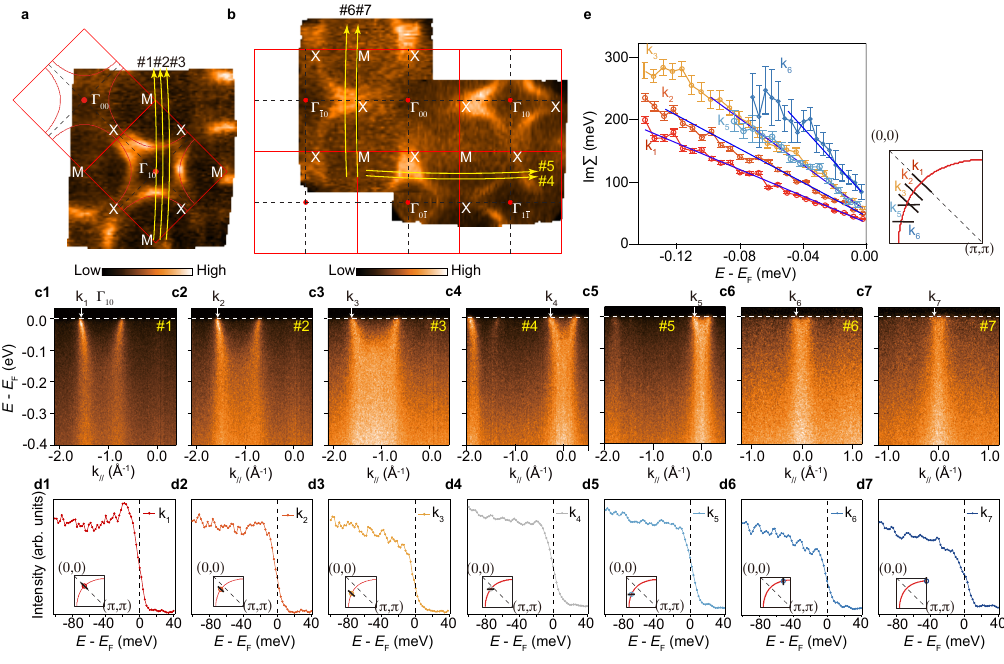}  
    \caption{\textbf{Momentum-dependent quasiparticle peak without pseudogap opening in La$_{0.8}$Ca$_{0.2}$NiO$_2$.}
\textbf{a,b}, Photoemission intensity maps in the $\Gamma$--$M$--$X$ plane, integrated over an energy window of $\pm 50$~meV around $E_{\mathrm{F}}$. The momentum locations of cut \#1–cut \#7 are indicated by yellow arrows.
\textbf{c1-c7}, ARPES spectra measured along cut \#1–cut \#7 in the $\Gamma$--$M$--$X$ plane, with the Fermi crossings of the $\alpha$ band labeled by white arrows as $k_1$–$k_7$.
\textbf{d1-d7}, Energy distribution curves (EDCs) at the corresponding Fermi momenta $k_F$, extracted from panels (\textbf{c1--c7}).
\textbf{e}, Imaginary part of the electron self-energy, $\mathrm{Im}\Sigma$, extracted from spectral analysis along cuts with Fermi-crossing momenta $k_1$, $k_2$, $k_3$, $k_5$, and $k_6$, corresponding to momenta moving from the $(\pi/2,\pi/2)$ direction toward $(\pi,0)$.
}

    \label{fig:figure2}
\end{figure*}

\begin{figure}[htbp]
    \centering
    \includegraphics[width=86mm]{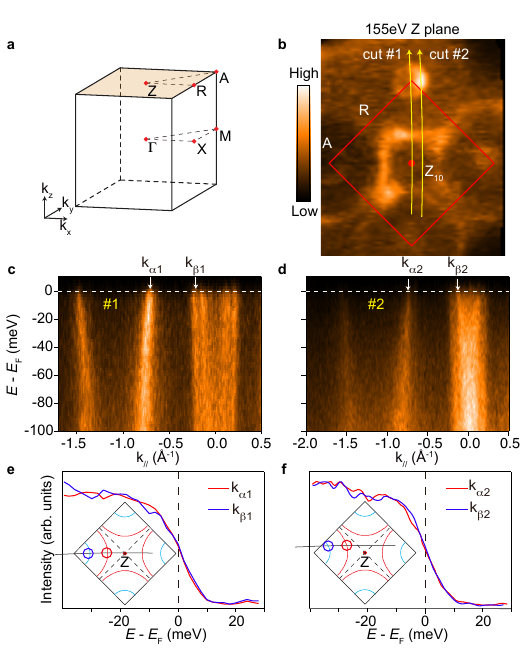}  
    \caption{\textbf{Absence of pseudogap opening in the $\beta$ band in La$_{0.8}$Ca$_{0.2}$NiO$_2$.}
\textbf{a}, Three-dimensional Brillouin zone (BZ) and high-symmetry points of La$_{0.8}$Ca$_{0.2}$NiO$_2$.
\textbf{b}, Photoemission intensity map integrated over an energy window of $\pm 50$~meV around $E_{\mathrm{F}}$ in the $Z$--$A$--$R$ plane. The momentum locations of cut \#1 and cut \#2 are indicated by yellow arrows.
\textbf{c,d}, Momentum-resolved ARPES intensity maps measured along cut \#1 (along $A$--$Z$--$A$) and cut \#2 (off $A$--$Z$--$A$), respectively.
\textbf{e,f}, Comparison of the EDCs at the Fermi momenta of the $\alpha$ and $\beta$ bands, as indicated by the arrows in arrows in panels (c) and (d), respectively. The spectra are normalized to allow direct comparison of the leading edges.
}

    \label{fig:figure1}
\end{figure}
\vspace{1em}
\noindent\textbf{Momentum-dependent spectral function of La$_{0.8}$Ca$_{0.2}$NiO$_2$}

As shown in Figs.~2a and 2b, Fermi surface maps measured with different in-plane sample orientations and across multiple Brillouin zones reveal a strong momentum-dependent modulation of spectral weight. 
While the spectral weight appears suppressed around certain X points, other symmetry-equivalent X points show strong intensity, suggesting that the intensity modulation primarily arises from photoemission matrix-element effect rather than Fermi-arc behavior.
We therefore focus on spectra at symmetry-equivalent momenta (cuts $\#1\sim\#7$ in Figs. 2a-b) where the spectral weight is maximal to examine the signature of pseudogap behavior.
The $\alpha$ band crosses the Fermi level in all momentum cuts as marked by white arrows $k_1$-$k_7$ (Figs. 2c1-c7).
The EDCs at Fermi momenta $k_1$-$k_7$ 
(Figs.~2d1-d7) 
show that the leading edge remains unchanged and fixed at the Fermi energy for all momenta. Even near $(\pi,0)$ (Fig.~2d7), a finite Fermi-edge drop is preserved. These observations indicate 
the absence of gap formation above the superconducting transition temperature in optimally doped La$_{0.8}$Ca$_{0.2}$NiO$_2$.

By extracting the imaginary part of the electron self-energy (Im$\Sigma$) from the spectra (Fig.~2e, see Supplementary Note III for details), we find that Im$\Sigma$ exhibits an approximately linear dependence on energy (Fig.~2e), deviating from the quadratic energy dependence expected for a Fermi liquid. This linear scaling over a 150 meV binding energy range is reminiscent of the marginal-Fermi-liquid behavior observed in cuprates \cite{VallaScience1999,VarmaPRL1989}. Intriguingly, its value at the Fermi level and the energy-dependent slope both increase monotonically toward $(\pi,0)$, indicating a strong reduction of the quasiparticle lifetime, enhanced inelastic scattering, and stronger electronic correlations from the momentum region near ($\pi$/2, $\pi$/2) towards ($\pi$, 0).

\begin{figure*}[t]
    \centering
    \includegraphics[width=170mm]{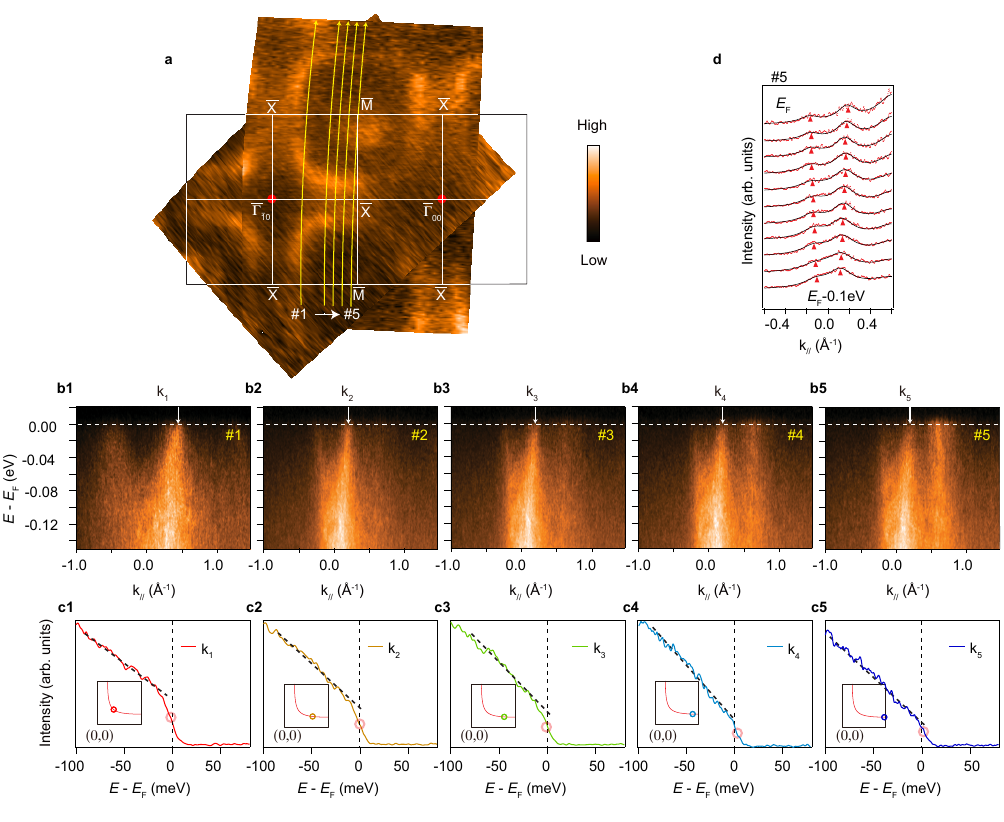}  
   \caption{\textbf{Momentum-dependent spectral-weight suppression without pseudogap opening in LaNiO$_2$.}
\textbf{a} Photoemission intensity map in the $\overline{\Gamma}$--$\overline{M}$--$\overline{X}$ plane, integrated over an energy window of $\pm 50$~meV around $E_{\mathrm{F}}$. The momentum locations of cut \#1–cut \#5 are indicated by yellow arrows.
\textbf{b1-b5} Momentum-resolved ARPES intensity maps along cut \#1–cut \#5, with the Fermi crossings of the $\alpha$ band marked by white arrows as $k_1$–$k_5$.
\textbf{c1-c5} Energy distribution curves (EDCs) at the corresponding Fermi-crossing momenta $k_1$–$k_5$. The slanted dashed lines indicate linear extrapolations of the leading-edge onset, and the open red circles mark the midpoint of the leading edge used to evaluate the possible leading-edge shift.
\textbf{d} Momentum distribution curves (MDCs) of cut \#5.
}

    \label{fig:figure3}
\end{figure*}

We further study the low-energy spectra of the $\beta$ band for possible signatures of a pseudogap.
High-resolution measurements taken along and off the A–Z–A direction (Figs.~3c-d) show that the $\beta$ band crosses the Fermi level along both the Z–A and A–R directions.
A comparison of the leading edges of the two EDCs at the Fermi momenta of the $\alpha$ and $\beta$ bands (Figs. 3e-f) shows that both leading edges remain pinned at the Fermi level, indicating the absence of detectable pseudogap opening in the $\beta$ band.




\begin{figure*}[ht]
    \centering
    \includegraphics[width=170mm]{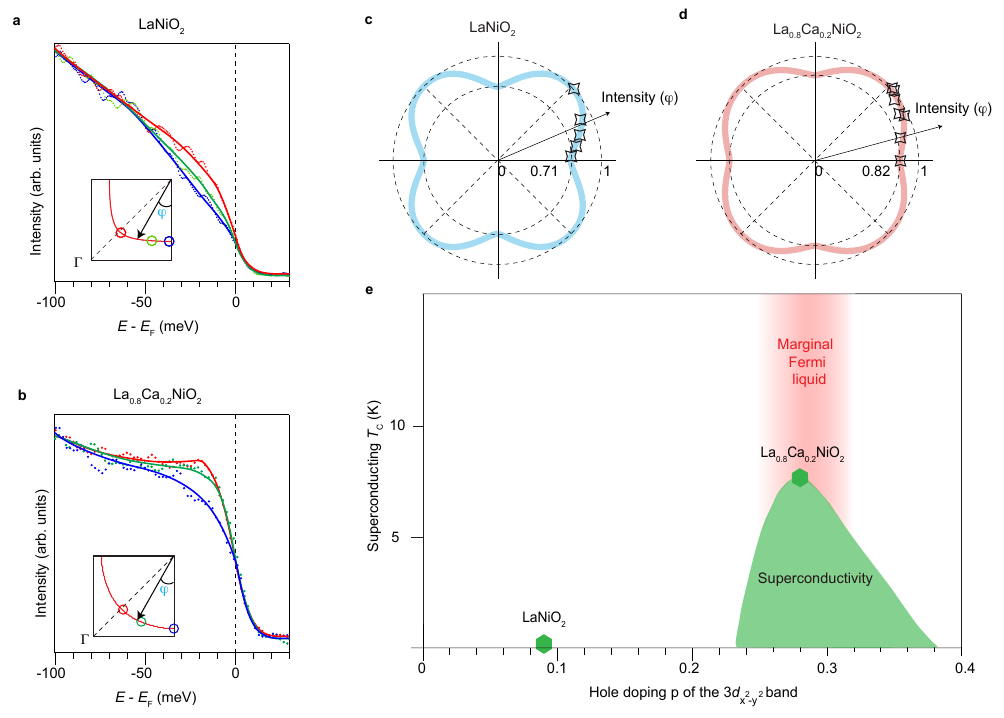}  
    \caption{\textbf{Anisotropic spectral weight suppression of La$_{1-x}$Ca$_{x}$NiO$_2$.}
\textbf{a,b}, Comparison of EDCs of the $\alpha$ band at different momenta for LaNiO$_2$ (\textbf{a}) and La$_{0.8}$Ca$_{0.2}$NiO$_2$ (\textbf{b}).
\textbf{c,d}, Momentum-dependent distribution of the density of states, integrated over the energy window [$E_{\mathrm{F}}$-50~meV, $E_{\mathrm{F}}$+10~meV], shown in polar coordinates for LaNiO$_2$ (\textbf{c}) and La$_{0.8}$Ca$_{0.2}$NiO$_2$ (\textbf{d}). The corresponding raw data are shown in Supplementary Fig.~S8. The radial coordinate represents the integrated spectral weight, and the polar angle represents $\phi$.
\textbf{e}, Our observation summarized in the phase diagram of La$_{1-x}$Ca$_x$NiO$_2$. 
Green markers denote the superconducting transition temperature. The effective hole doping $p$ of the Ni $d_{x^2-y^2}$ band is taken from ref.~\onlinecite{dingNSR2024}.}

    \label{fig:figure4}
\end{figure*}

\vspace{1em}
\noindent\textbf{Momentum-dependent spectral function of LaNiO$_2$}

To understand the doping dependence of the quasiparticle dispersion and examine whether pseudogap behavior emerges at lower doping, we further study the low-energy spectra of LaNiO$_2$. Its effective hole concentration in the $d_{x^2-y^2}$ band has been estimated to be $p\sim 0.09$ (ref.~\onlinecite{dingNSR2024}), placing it in a doping regime where the pseudogap is well developed in hole-doped cuprates~\cite{Kivelsonnature2013}.

Photoemission intensity maps acquired with 102 eV photons show that the spectral weight near the ($\pi,0$) region is not strongly suppressed by matrix-element effects (Fig.~4a, see Supplementary Fig.~S6 for details). Owing to the quasi-two-dimensional nature of the $\alpha$ band \cite{dingNSR2024}, slight deviations from the $\Gamma$–X–M plane should not qualitatively alter its dispersion.
Around ($\pi/2, \pi$/2), the density of states near the Fermi level relative to high binding energy is much weaker
in LaNiO$_2$ than in La$_{0.8}$Ca$_{0.2}$NiO$_2$ (Figs. 4b1 and c1).
Nevertheless, as shown in Figs.~4b1–b5, the $\alpha$ band approaches and crosses the Fermi level at all measured momenta (labeled $k_1$–$k_5$).
This is further confirmed by the MDCs around $(\pi,0)$(Fig.~4d), revealing continuous dispersion of $\alpha$ band up to the Fermi level, in clear contrast to the back-bending behavior characteristic of the pseudogap state in cuprates \cite{DamascelliRMP2003}.
The associated EDCs retain finite spectral weight across all momentum cuts, including those near $(\pi,0)$ (Figs.~4c1-c5), where cuprates at a similar hole doping show a leading edge gap over tens of meV due to the pseudogap behavior \cite{KondoNature2009}.
\par


\vspace{1em}

\noindent\textbf{Implications for the nickelate phase diagram}

Within our experimental sensitivity and momentum coverage, both LaNiO$_2$ and La
$_{0.8}$Ca$_{0.2}$NiO$_2$ exhibit gapless Fermi crossings around the measured Fermi surfaces, suggesting continuous large Fermi pockets rather than Fermi arcs.
 In hole-doped cuprates, by contrast, a pseudogap opens well above $T_c$ and is typically substantially larger than the superconducting gap. Even in the single-layer cuprates with the lowest maximal $T_c$, such as La$_{2-x}$Sr$_x$CuO$_4$, the antinodal pseudogap remains of order 10–20 meV near optimal doping and increases to 30–40 meV around $p\sim0.10$ (ref.~\onlinecite{YoshidaPRB2016}). In Bi$_2$Sr$_{2-x}$La$_x$CuO$_{6+\delta}$, the pseudogap reaches approximately 20–40 meV near optimal doping and 40–50 meV at $p\sim0.10$ (refs.~\onlinecite{KondoNature2009,HashimotoNatPhys2010}). These energy scales are significantly larger than our experimental gap-detection limit of $\sim$2 meV, which is known to be approximately one-fifth of the experimental energy resolution in ARPES experiments~\cite{ZhangNatPhys2012}, indicating that a comparable gap would be readily resolved. Consistently, neither the $\alpha$ nor the $\beta$ band exhibits back-bending, and the leading edge remains pinned at $E_F$. These results indicate that, within the doping range examined, infinite-layer nickelates do not exhibit a detectable pseudogap regime of the type observed in hole-doped cuprates.

Despite the absence of a pseudogap, the cuprate-like $\alpha$ band exhibits pronounced momentum-dependent spectral-weight suppression. As shown in Figs.~5a and 5b, both LaNiO$_2$ and La$_{0.8}$Ca$_{0.2}$NiO$_2$ display a progressive reduction of low-energy spectral weight upon approaching $(\pi,0)$. The corresponding polar plots (Figs.~5c,d) reveal a continuous decrease in photoemission intensity within the energy window [$E_{\mathrm{F}}$-50~meV, $E_{\mathrm{F}}$+10~meV] as the polar angle $\phi$ increases from $0^\circ$ to $45^\circ$. The suppression is more pronounced in LaNiO$_2$: the spectral weight at $(\pi,0)$ ($\phi=45^\circ$) is reduced to 0.71 of that near $(\pi/2,\pi/2)$, compared with 0.81 in La$_{0.8}$Ca$_{0.2}$NiO$_2$.
Together with the observed marginal-Fermi-liquid-like self-energy (Fig. 2e), these results reveal an unusual phase diagram in which momentum-dependent correlation effects and marginal-Fermi-liquid behavior emerge without an accompanying pseudogap regime. 
\vspace{1em}
\noindent\textbf{Discussion and Outlook}


Pseudogap is widely regarded as a defining feature of the normal state from which superconductivity emerges in hole-doped cuprates. The origin of the pseudogap in cuprates remains debated, with interpretations ranging from preformed Cooper pairing \cite{KondoNatPhys2011} to competing ordered states like density wave order \cite{CominScience2014,HashimotoNatMater2015}. 
Our observation of gapless low-energy spectra in La$_{1-x}$Ca$_x$NiO$_2$ shows that superconducting infinite-layer nickelates do not require a detectable cuprate-like pseudogap in the normal-state single-particle spectrum.
In addition, charge-density-wave order, which is frequently discussed in connection with the cuprate pseudogap \cite{CominScience2014}, has not been established in infinite-layer nickelates \cite{parzyckNM2024}. 
These results indicate that hole doping of a $3d^9$-derived electronic structure, as realized in infinite-layer nickelates, is insufficient by itself to generate a cuprate-like pseudogap.

The lack of a cuprate-like pseudogap in infinite-layer nickelates may be related to their distinct charge-transfer energetics. Hole-doped cuprates are charge-transfer systems in which doped holes have substantial O \(2p\) character and form Zhang--Rice-like states \cite{ChenPRL1991,ZhangPRB1988}. In contrast, electron-doped cuprates populate more Cu \(3d\)-derived states on the opposite side of the charge-transfer gap, and do not generally exhibit the broad antinodal pseudogap phenomenology characteristic of underdoped hole-doped cuprates \cite{ArmitageRMP2010,HorioNatCommun2016}.
Infinite-layer nickelates appear to represent a further distinct limit  closer to the Mott--Hubbard regime, with doped holes residing predominantly on Ni \(3d\) orbitals rather than ligand oxygen states \cite{GoodgePNAS2021,ChenPRL2022}. Together with recent three-orbital calculations\cite{PengPRB2025} showing that pseudogap-like spectral features are suppressed as the charge-transfer energy increases from the cuprate to the nickelate regime, this comparison suggests that charge-transfer energetics and doped-carrier orbital character are important parameters controlling whether momentum-dependent correlations evolve into a cuprate-like pseudogap.


Despite the absence of a pseudogap, La$_{1-x}$Ca$_x$NiO$_2$ shares several notable spectroscopic characteristics with cuprates, including marginal-Fermi-liquid-like scattering and pronounced momentum-dependent correlations. 
In the Ni $d_{x^2-y^2}$ band, the low-energy spectral features near \((\pi,0)\) are strongly broadened and reduced while the underlying dispersion remains gapless. This behavior is reminiscent of the non-Fermi-liquid quasiparticle dynamics observed in slightly overdoped La$_{1.77}$Sr$_{0.23}$CuO$_4$ (ref.~\onlinecite{ChangNatCommun2013}).
The effect becomes more pronounced upon approaching the parent compound. Although subtle differences in residual disorder cannot be completely ruled out, both LaNiO$_2$ and La$_{0.8}$Ca$_{0.2}$NiO$_2$ films were prepared under similarly optimized conditions, suggesting that the enhanced anisotropy in LaNiO$_2$ is at least partly intrinsic and associated with reduced hole doping and stronger electronic correlations.
This trend is consistent with the observation of short-range magnetic correlations and spin-glass behavior reported on the low-doping side of the nickelate phase diagram \cite{FowlieNatPhys2022,OrtizPRR2022,SaykinnpjQM2025}. It may also provide a single-particle spectral interpretation for the suppression of the Knight shift reported in polycrystalline LaNiO$_2$ \cite{zhaoPRL2021}.

The self-energy analysis provides further evidence for the momentum dependence of electronic correlations. In La$_{0.8}$Ca$_{0.2}$NiO$_2$, Im\(\Sigma(\omega)\) exhibits an approximately linear energy dependence, consistent with marginal-Fermi-liquid-like phenomenology within our experimental resolution. More importantly, the slope of Im\(\Sigma(\omega)\) increases toward \((\pi,0)\), indicating that quasiparticle damping is strongly momentum dependent. 
These observations show that momentum-dependent correlations and non-Fermi-liquid-like scattering can persist even when no cuprate-like pseudogap is detected. They also highlight a similarity to cuprates, where electronic correlations are generally strongest in the antinodal region and become more pronounced upon entering the underdoped regime. 


An important open question is the microscopic origin of the enhanced correlations near $(\pi,0)$. Possible ingredients include spin fluctuations and multiorbital correlations, calling for further theoretical development. Infinite-layer nickelates reproduce several central aspects of cuprate phenomenology, including a cuprate-like $d_{x^2-y^2}$ Fermi surface, superconductivity, strange-metal-like scattering and momentum-dependent quasiparticle damping, while lacking a detectable cuprate-like pseudogap in the doping range examined here. This contrast provides a useful benchmark for identifying the electronic parameters that should be incorporated into theories of high-temperature superconductivity.

\par
\par
\noindent\textbf{Acknowledgments}

We thank F. C. Zhang and K. Jiang for helpful discussions and Z. T. Liu for experimental support. This work was carried out with the support of Diamond Light Source, instrument I05 (proposal SI39544) and BL03U of the Shanghai Synchrotron Radiation Facility. This work was supported by the National Natural Science Foundation of China (Grants Nos. 92477206, 12422404, 12274085, 92365302), the National Key R\&D Program of China (2023YFA1406300), the New Cornerstone Science Foundation, the Quantum Science and Technology-National Science and Technology Major Project (2021ZD0302803) and the Shanghai Municipal Science and Technology Major Project (2019SHZDZX01).

\normalem

\end{document}